\documentclass[]{emulateapj}
\usepackage{graphicx,amsmath}

\begin{document}
\title{Asymptotic self-similar solutions with a characteristic time-scale}

\author{Eli Waxman\altaffilmark{1} and Dov Shvarts\altaffilmark{2,3}}
\altaffiltext{1}{Department of Particle Physics \& Astrophysics, Weizmann Institute of Science, Rehovot 76100, Israel}
\altaffiltext{2}{Department of Physics, Nuclear Research Center Negev, P.O.B. 9001, Beer-Sheva 84015, Israel}
\altaffiltext{3}{Department of Mechanical Engineering, Ben-Gurion University of the Negev, Be'er-Sheva, Israel}
\date{\today}

\begin{abstract}

For a wide variety of initial and boundary conditions, adiabatic one dimensional flows of an ideal gas approach self-similar behavior when the characteristic length scale over which the flow takes place, $R$, diverges or tends to zero. It is commonly assumed that self-similarity is approached since in the $R\rightarrow\infty(0)$ limit the flow becomes independent of any characteristic length or time scales. In this case, the flow fields $f(r,t)$ must be of the form $f(r,t)=t^{\alpha_f}F(r/R)$ with $R\propto(\pm t)^\alpha$. We show that requiring the asymptotic flow to be independent only of characteristic length scales implies a more general form of self-similar solutions, $f(r,t)=R^{\delta_f}F(r/R)$ with $\dot{R}\propto R^\delta$, which includes the exponential ($\delta=1$) solutions, $R\propto e^{t/\tau}$. We demonstrate that the latter, less restrictive, requirement is the physically relevant one by showing that the asymptotic behavior of accelerating blast-waves, driven by the release of energy at the center of a cold gas sphere of initial density $\rho\propto r^{-\omega}$, changes its character at large $\omega$: The flow is described by $0\le\delta<1$, $R\propto t^{1/(1-\delta)}$, solutions for $\omega<\omega_c$, by $\delta>1$ solutions with $R\propto (-t)^{1/(\delta-1)}$ diverging at finite time ($t=0$) for $\omega>\omega_c$, and by exponential solutions for $\omega=\omega_c$ ($\omega_c$ depends on the adiabatic index of the gas, $\omega_c\sim8$ for $4/3<\gamma<5/3$). The properties of the new solutions obtained here for $\omega\ge\omega_c$ are analyzed, and self-similar solutions  describing the $t>0$ behavior for $\omega>\omega_c$ are also derived.

\end{abstract}

\maketitle

\section{Introduction}
\label{sec:introduction}

Self-similar solutions to the hydrodynamic equations describing adiabatic one dimensional flows of an ideal gas are of interest for several reasons. The non-linear partial differential hydrodynamic equations are reduced for self-similar flows to ordinary differential equations, which greatly simplifies the mathematical problem of solving the equations and in certain cases allows one to find analytic solutions. Moreover, self-similar solutions often describe the limiting behavior approached asymptotically by flows which take place over a characteristic scale, $R$, which diverges or tends to zero \citep[see][for reviews]{SedovBook,ZelDovichRaizer,BarenblattBook}. Some examples of such asymptotic solutions which are widely used in astrophysical contexts are the Sedov-von Neumann-Taylor solutions  \citep{Sedov46,vonNeumann47,Taylor50} describing expanding decelerating spherical blast waves, for which $R\rightarrow\infty$, and the Gandel'Man-Frank-Kamenetskii--Sakurai solutions \citep{GandelMan56,Sakurai60} describing the emergence of a shock wave from the surface of a star, for which $R\rightarrow0$. Both types of solutions are relevant, e.g., to supernova explosions and in particular to the recently detected shock breakouts \citep[e.g.][]{MM99,WaxmanMeszarosCampana07}. An extensive discussion of spherical self-similar blast-waves in an astrophysical context is given by \citet{OstrikerMcKee88}.

It is commonly assumed that self-similarity is approached since in the $R\rightarrow\infty(0)$ limit the flow becomes independent of any characteristic length or time scales. Using dimensional arguments it is possible to show that if the flow is determined by a set of constants, using which it is impossible to construct a constant with the dimensions of length or time, then the flow fields $f(r,t)$ must be of the form $f(r,t)=t^{\alpha_f}F(r/R)$ with $R\propto(\pm t)^\alpha$ \citep[see chapter XII of][]{ZelDovichRaizer}. We show in \S~\ref{sec:general} that requiring the asymptotic flow to be independent only of characteristic length scales is sufficient for showing, based on dimensional arguments, that the flow must be self-similar. The less restrictive requirement allows a more general form of self-similar solutions, $f(r,t)=R^{\delta_f}F(r/R)$ with $\dot{R}\propto R^\delta$, which includes the exponential ($\delta=1$) solutions, $R\propto e^{t/\tau}$. The existence of exponential self-similar solutions has been noted by several authors \citep[e.g.][]{StanyukovichBook}. However, it was generally assumed that asymptotic solutions, which are of interest, are of a power law form.

In \S~\ref{sec:blast} we show that the asymptotic self-similar solutions describing the propagation of accelerating blast waves, propagating in a cold gas sphere of initial density $\rho\propto r^{-\omega}$ with $\omega>3$, are of the more general form, $\dot{R}\propto R^\delta$, with exponential solutions obtained at $\omega=\omega_c(\gamma)$, where $\gamma$ is the adiabatic index of the gas. The new solutions obtained here for $\omega\ge\omega_c$ extend the family of second type solutions describing the asymptotic flow of accelerating blast waves, which was derived by \citet{WaxmanShvarts93} and was limited to $\omega<\omega_c$, to $\omega\ge\omega_c$. The properties of the new solutions are analyzed in \S~\ref{sec:WS}, and self similar solutions describing the $\omega>\omega_c$ flow at times later than the finite divergence time are derived in \S~\ref{sec:t0}. Our results are summarized in \S~\ref{sec:summary}.

It should be noted here that the self-similar solutions derived by \citet{WaxmanShvarts93} exist only for $\omega>\omega_g(\gamma)>3$, where $\omega_g=3.26$ for $\gamma=5/3$ and approaches 3 for $\gamma\rightarrow1$, while the Sedov-von Neumann-Taylor solutions provide the correct asymptotic solutions only for $\omega<3$. The asymptotic behavior within the (narrow) range of $3<\omega<\omega_g(\gamma)$ is not described by either of the two types of solutions. The nature of the asymptotic flow in this regime is discussed in \citet{Gruzinov03,KushnirGap}.

\section{The general form of asymptotic self-similar solutions}
\label{sec:general}

The equations describing adiabatic one-dimensional flow of an ideal gas are \citep[e.g.][]{ZelDovichRaizer}
\begin{eqnarray}
\label{eq:hydro_eq}
(\partial_{t}+u\partial_{r})\ln\rho+ r^{-(\nu-1)}\partial_{r}(r^{\nu-1}u) &=& 0,
\nonumber \\
(\partial_{t}+u\partial_{r})u+\rho^{-1}\partial_{r}(\gamma^{-1}\rho c^{2}) &=&
0, \nonumber \\
(\partial_{t}+u\partial_{r})(c^{2}\rho^{1-\gamma}) &=& 0,
\end{eqnarray}
where $u$, $c$, and $\rho$ are the fluid velocity, sound speed and density respectively (the pressure is
given by $p=\rho c^{2}/\gamma$), and $\nu=1,2,3$ for planar, cylindrical and spherical symmetry respectively.

Consider a solution $f(r,t;\{c_i\},\gamma)$, where $f$ stands for $\rho$, $u$ or $c$, and $\{c_i\}$ are the set of constants determining the initial and boundary conditions. Let us assume that the flow takes place over a characteristic scale $R(t)$, which diverges or tends to zero monotonically with time (if $R(t)$ is not monotonic then the flow is characterized by some finite length scales corresponding to the extrema of $R$). In this case we may replace the variables $\{r,t\}$ with $\{\xi(r,t)=r/R,R\}$, and describe the flow by $\tilde{f}(\xi,R;\{c_i\},\gamma)$, where $f(r,t;\{c_i\},\gamma)=\tilde{f}[\xi(r,t),R(t);\{c_i\},\gamma]$, and by $\dot{R}(R;\{c_i\},\gamma)$, which determines the relation between $R$ and $t$.

Let us assume next that in the limit $R\rightarrow\infty(0)$ the flow becomes independent of any characteristic length scales. That is, that in the limit $R\rightarrow\infty(0)$ the flow is determined by only a subset of $\{c_i\}$, denoted $\{b_i\}$, from which a constant with the dimensions of length cannot be constructed. The dimensions of each parameter $b_k$ may be considered as a three dimensional vector $\vec{v}_k$, $[b_k]=M^{v_1}L^{v_2}T^{v_3}$, where $M,\,L$, and $T$ are the units of mass, length, and time, respectively. Since a constant with the dimensions of length may not be constructed using $\{b_i\}$, there are at most two constants with independent dimension vectors. As we shall see below, there must exit two constants with independent dimensions. Denoting these by $\{a_1,a_2\}$, and expressing the other constants as $b_i=\gamma_i a_1^{v_{i1}} a_2^{v_{i2}}$ with dimensionless $\gamma_i$, the asymptotic solution may be written as $\tilde{f}_{\rm as.}(\xi,R;a_1,a_2,\{\gamma_i\},\gamma)$, $\dot{R}_{\rm as.}(R;a_1,a_2,\{\gamma_i\},\gamma)$. Since the dimensions of $\{R,a_1,a_2\}$ are independent (otherwise it would be possible to construct a constant with dimensions of length using $\{a_1,a_2\}$), there is a single combination (product of powers) of $\{R,a_1,a_2\}$ that has a given dimension vector. This implies that $\tilde{f}_{\rm as.}$ must be of the form $\tilde{f}_{\rm as.}=a_1^{\alpha_{f1}}a_2^{\alpha_{f2}}R^{\alpha_{f}}F(\xi;\{\gamma_i\},\gamma)$ and $\dot{R}_{\rm as.}$ must be of the form $\dot{R}_{\rm as.}=a_1^{\delta_{1}}a_2^{\delta_{2}}R^{\delta}F_R(\{\gamma_i\},\gamma)$.

Thus, the assumption that the asymptotic solution is independent of any characteristic length scales implies that it must be a self-similar solution of the form
\begin{equation}\label{eq:ss_scaling}
    u=\dot{R}\xi U(\xi),\quad c=\dot{R}\xi C(\xi),\quad \rho=BR^\epsilon G(\xi)
\end{equation}
with
\begin{equation}\label{eq:Rdot}
    \dot{R}=AR^\delta.
\end{equation}
The spatial part of $u$($c$) was chosen as $\xi U(\xi)$ ($\xi C(\xi)$) for convenience (see below). Note, that the dimensions of $A$ and $B$ are independent. These solutions include the $R\propto(\pm t)^\alpha$ solutions, obtained for $\delta=(\alpha-1)/\alpha\neq1$, and the exponential $R\propto e^{t/\tau}$ solutions obtained for $\delta=1$.

Substituting the ansatz of eqs.~(\ref{eq:ss_scaling}) and~(\ref{eq:Rdot}) in the hydrodynamic equations (eq.~(\ref{eq:hydro_eq})), the partial differential equations are replaced with a single ordinary differential equation,
\begin{equation}\label{eq:dUdC}
\frac{dU}{dC}=\frac{\Delta_{1}(U,C)}{\Delta_{2}(U,C)},
\end{equation}
and one quadrature
\begin{equation}\label{eq:quadrature}
\frac{d\ln\xi}{dU}=\frac{\Delta(U,C)}{\Delta_{1}(U,C)}\qquad {\rm
or} \qquad \frac{d\ln\xi}{dC}=\frac{\Delta(U,C)}{\Delta_{2}(U,C)}.
\end{equation}
$G$ is given implicitly by
\begin{equation}\label{eq:G}
(\xi C)^{-2(\nu+\epsilon)}|1-U|^{\lambda}G^{(\gamma-1)(\nu+\epsilon)+\lambda}\xi^{\nu\lambda}={\rm Const.},
\end{equation}
with
\begin{equation}\label{eq:lambda}
\lambda=-(\gamma-1)\epsilon+2\delta.
\end{equation}
The functions $\Delta$, $\Delta_{1}$, and $\Delta_{2}$ are
\begin{eqnarray}\label{eq:deltas}
\Delta&=&C^{2}-(1-U)^{2}, \nonumber \\
\Delta_{1}&=&U(1-U)(1-U-\delta)-C^{2}\left(\nu U+\frac{\epsilon+2\delta}{\gamma}\right), \nonumber \\
\Delta_{2}&=&C\{(1-U)(1-U-\delta) \nonumber \\
&-&\frac{\gamma-1}{2}U\left[(\nu-1)(1-U)+\delta\right]-C^{2} \nonumber \\
&+&\frac{2\delta-(\gamma-1)\epsilon}{2\gamma}\frac{C^{2}}{1-U}\}.
\end{eqnarray}

\section{Accelerating blast wave solutions}
\label{sec:blast}

Consider the blast wave produced by the deposition of energy $E$ within a region of characteristic size $d$ at the center of an initially cold ($p=0$ at $r>d$) gas sphere with initial density $\rho_0=K r^{-\omega}$ (at $r>d$). As the shock radius $R$ diverges, it is reasonable to assume that the flow becomes independent of any length scale determined by the initial conditions, i.e. that $R(t)$ is the only relevant length scale. In this case, the flow should approach a self-similar solution of the form given by eqs.~(\ref{eq:ss_scaling}) and~(\ref{eq:Rdot}). Since the density just behind the shock wave is, for strong shocks, a constant factor, $(\gamma+1)/(\gamma-1)$, times the density just ahead of the shock, we must have $\epsilon=-\omega$, and we may choose $B=K$. With this normalization, the Rankine-Hugoniot relations at the shock front determine the boundary conditions for the self-similar solutions to be \citep[e.g.][]{ZelDovichRaizer}
\begin{equation}\label{eq:shock_boundary}
    U(1)=\frac{2}{\gamma+1},\quad C(1)=\frac{\sqrt{2\gamma(\gamma-1)}}{\gamma+1}, \quad G(1)=\frac{\gamma+1}{\gamma-1}.
\end{equation}
The only parameter of the self-similar solution that remains to be determined is $\delta$. The methods described below for determining $\delta$, and for analyzing the resulting solutions' properties, are similar to those described in \citep{WaxmanShvarts93}. The discussion in the following subsections is therefore very concise. The reader may refer to \cite{WaxmanShvarts93} for more elaborate explanations.

\subsection{$\omega<\omega_c$}
\label{sec:WS}

In the Sedov-von Neumann-Taylor analysis it is assumed that the second dimensional constant, in addition to $K$, that determines the self-similar solution is $E$. In this case dimensional considerations imply $R\propto (Et^2/K)^{1/(5-\omega)}$, i.e. $\delta=(\omega-3)/2$. As explained in detail in \citep{WaxmanShvarts93}, the Sedov-von Neumann-Taylor solutions are the correct asymptotic solutions only for $\omega<3$, for which $\delta<0$ and the blast wave decelerates with time. For larger values of $\omega$ the mass and energy contained within the self-similar solution are infinite, reflecting the fact that the initial gas mass at $r>d$ diverges for $d\rightarrow0$. It was therefore suggested by \citet{WaxmanShvarts93} that for $\omega>3$ the asymptotic solution is given by a self-similar solution only over part of the $\{\xi,R\}$ plane, bounded by $\xi=1$ and $\xi_c(R)<1$, and by a different solution at $0<\xi<\xi_c(R)$.

Since $\xi_c(R)$ must be a contact or a weak discontinuity, $\xi_c(R)$ must be a characteristic of the self-similar solution. For the self-similar flow, the characteristic lines
\begin{equation}\label{eq:characteristics}
    C_0:\frac{dr_0}{dt}=u,\quad C_\pm: \frac{dr_\pm}{dt}=u\pm c,
\end{equation}
are given by
\begin{eqnarray}\label{eq:slfsim_char}
C_{0}&:&\frac{d \ln \xi_{0}}{d \ln R}=U(\xi_{0})-1,\nonumber \\
C_{\pm}&:&\frac{d \ln \xi_{\pm}}{d \ln R}=U(\xi_{\pm})\pm
C(\xi_{\pm})-1.
\end{eqnarray}
The flow just behind the shock is always subsonic: the shock-front point $\{U=U(1),C=C(1)\}$ lies above the "sonic-line" $U+C=1$, which implies that $C_+$ characteristics emerging from points just behind the shock always overtake it. $C_+$ characteristics that do not overtake the shock exit only if the self-similar solution crosses the $U+C=1$ line in the $\{U,C\}$ plane into the region where $U+C<1$. Requiring $\xi_c$ to coincide with a $C_+$ characteristic, that does not overtake the shock, implies therefore that the solution must cross the $U+C=1$ line. Since $\Delta=0$ along the sonic line, eqs.~(\ref{eq:quadrature}) imply that a physical solution must cross the sonic line at a singular point $\Delta_1=\Delta_2=0$ (otherwise $U(\xi)$ or $C(\xi)$ are not single valued). This requirement determines the correct value of $\delta$ for the $\omega>3$ asymptotic solutions.

The self-similar solutions obtained in this way for $\omega>3$ were analyzed in detail in \citep{WaxmanShvarts93}. They describe accelerating blast waves, with $\delta>0$, and approach the singular point $\{U=1-\delta,C=0\}$ as $\xi\rightarrow0$. Analyzing the behavior of the solutions near this singular point, it was shown that although the mass and energy contained in the self-similar solution are infinite, the mass and energy contained within the region $\xi_c(R)<\xi<1$, where $\xi_c(R)$ is a $C_+$ characteristic which satisfies $\xi_c(R)\rightarrow0$ as $R\rightarrow\infty$, approach finite values as $R\rightarrow\infty$. Moreover, it was shown that $\xi_c(R)R\propto t$ as $R\rightarrow\infty$, implying that the asymptotic flow within the region $0<r<\xi_c(R)R$ is described by the self-similar solution of expansion into vacuum. Finally, it was demonstrated in  \citep{WaxmanShvarts93} by numerical simulations that the asymptotic behavior described above is indeed approached for $R/d\gg1$.

\subsection{$\omega\ge\omega_c$}
\label{sec:WS}

\begin{figure}
\includegraphics[scale=0.44]{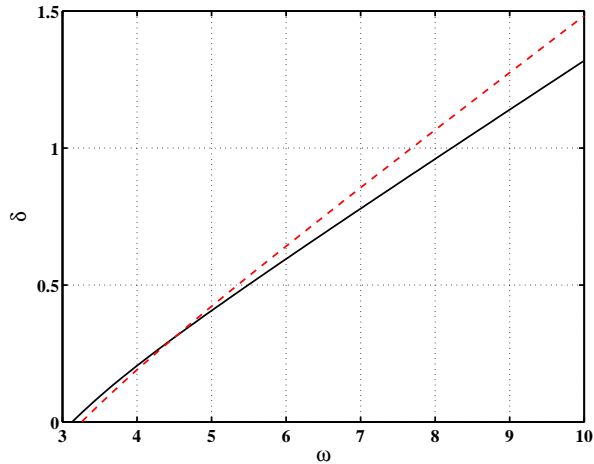}
\caption{$\delta(\omega)$ for $\gamma=5/3$ (red dashed line) and $\gamma=4/3$ (solid black line) determined by the requirement that the $C(U)$ curve crosses the sonic line at a singular point. As noted in \citep{WaxmanShvarts93}, there are no solutions satisfying this requirement within some narrow range $3<\omega<\omega_g(\gamma)$ ($\omega_g=3.26$ for $\gamma=5/3$ and approaches 3 for $\gamma\rightarrow1$). The nature of the asymptotic flow in this regime is discussed in \citep{KushnirGap}.}
\label{fig:delta}
\end{figure}

\begin{figure}
\includegraphics[scale=0.44]{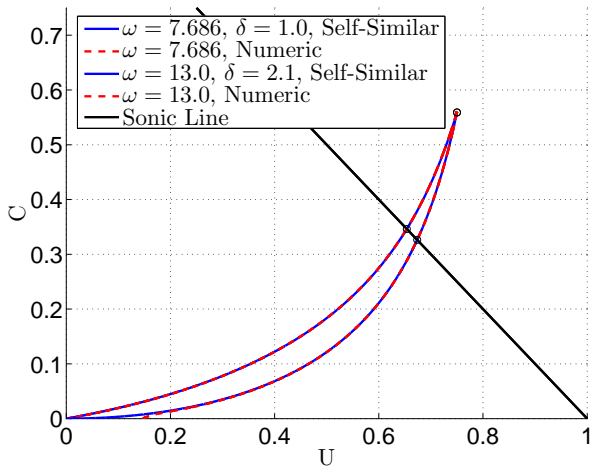}
\caption{A comparison of the $C(U)$ curves obtained from numerical solutions of the hydrodynamic equations, eq.~(\ref{eq:hydro_eq}), with those obtained by solving the self-similar eq.~(\ref{eq:dUdC}). Results are shown for $\gamma=5/3$ and two values of $\omega$, $\omega=7.686$ for which the self-similar analysis yields $\delta=1.0$ (lower curve), and $\omega=13.0$, for which the self-similar analysis gives $\delta=2.10$. Numerical solutions are shown for $R/d=500$ (where $R$ is the shock radius and $d$ the radius of the sphere within which the energy is initially deposited).}
\label{fig:UC}
\end{figure}

In the analysis of \citep{WaxmanShvarts93} it was assumed that the asymptotic solutions are of the form $R\propto t^\alpha$. It was found that $\alpha$ diverges as $\omega$ approaches a finite value, $\omega_c(\gamma)$ \citep[$\omega_c\sim8$ for $4/3<\gamma<5/3$ and diverges for $\gamma\rightarrow1$, see fig.~5 of][]{WaxmanShvarts93}.
Allowing solutions of the more general form, $\dot{R}\propto R^\delta$ as suggested in \S~\ref{sec:general}, the divergence of $\alpha$ suggests that the asymptotic solution for $\omega=\omega_c(\gamma)$ is an exponential solution, $R\propto e^{t/\tau}$ (recall that $\delta=(\alpha-1)/\alpha$). Searching for solutions of the form $\dot{R}\propto R^\delta$, that cross the sonic line at a singular point, we find that such solutions do exist for $\omega\ge\omega_c$.  $\delta(\omega)$ is shown for $\gamma=5/3$ and $\gamma=4/3$ in figure~\ref{fig:delta}. We find $\delta=1$ for $\omega=\omega_c(\gamma)$, implying exponential expansion, and $\delta>1$ for $\omega>\omega_c(\gamma)$, implying that the shock radius diverges in finite time, $R\propto (-t)^{1/(1-\delta)}$. Fig.~\ref{fig:UC} presents a comparison of the $C(U)$ curves obtained by numerically solving the hydrodynamic equations, eq.~(\ref{eq:hydro_eq}), with those obtained by solving the self-similar eq.~(\ref{eq:dUdC}). The numerical results suggest that the self-similar solutions provide a correct description of the asymptotic flow.

For $\delta\ge1$, the self-similar solution approaches the singular point $\{U=0,C=0\}$ as $\xi\rightarrow0$. Solving the self-similar equations in the vicinity of this point we find that in the limit of $\xi\rightarrow0$ the solutions are given by
\begin{equation}\label{eq:asym_d1}
    U=-\frac{1}{\ln\xi},\quad C\propto U^{(3\gamma-1)/2},\quad G\propto\xi^{-\omega}(\ln\xi)^{\omega-3}
\end{equation}
for $\delta=1$, and by
\begin{equation}\label{eq:asym_dg1}
    U=U_1\xi^{\delta-1},\quad C=C_1 \xi^{\delta-1},\quad G=G_1\xi^{-\omega},
\end{equation}
where $U_1,C_1,G_1$ are constants, for $\delta>1$. Using eqs.~(\ref{eq:asym_d1}) and~(\ref{eq:asym_dg1}) we find that the $C_+$ characteristics originating from points below the sonic line are given in the limit $R\rightarrow\infty$ by
\begin{equation}\label{eq:C+}
    \xi_+R\propto\left\{%
\begin{array}{ll}
    |\ln\xi_+|, & \hbox{$\delta=1$;} \\
    {\rm Const.}, & \hbox{$\delta>1$.} \\
\end{array}%
\right.
\end{equation}

The energy contained in a region of the self-similar solution, bounded by $\xi=1$ and a $C_+$ characteristic $\xi_+(R)$, is given by
\begin{eqnarray}\label{eq:E_ss}
    E_+(R)=&4&\pi K A^2 R^{3-\omega+2\delta} \nonumber\\
           &\times&\int_{\xi_+(R)}^1d\xi\xi^4G\left[\frac{1}{2}U^2+\frac{1}{\gamma(\gamma-1)}C^2\right].
\end{eqnarray}
In order for the energy $E_+$ not to diverge as $R\rightarrow\infty$ and $\xi_+(R)\rightarrow0$, we must have
\begin{equation}\label{eq:d_lim}
    \delta<\frac{\omega-3}{2}.
\end{equation}
This is satisfied for the solutions presented in fig.~\ref{fig:delta}. For $\delta$ values satisfying eq.~(\ref{eq:d_lim}), the integral over $\xi$ diverges as $\xi_+(R)\rightarrow0$, while $R^{3-\omega+2\delta}$ vanishes as $R\rightarrow\infty$. Using eqs.~(\ref{eq:asym_d1})--(\ref{eq:C+}), it is straight forward to verify that $E_+(R)\rightarrow{\rm Const.}$ as $R\rightarrow\infty$. Thus, the energy contained in the region of the flow which is described by the self-similar solution, $\xi_c(R)<\xi<1$, approaches a finite value as $R\rightarrow\infty$. A similar calculation shows that the mass contained in this region also approaches a finite value.

\subsection{$t>0$}
\label{sec:t0}

For $\delta>1$, the shock radius diverges at a finite time, $t=0$. This implies that the spatial distribution of the flow fields at the divergence time is described by the $\xi\rightarrow0$ behavior of the solution, which is given by eq.~(\ref{eq:asym_dg1}). That is, at $t=0$ the spatial distribution of the flow fields is
\begin{equation}\label{eq:t0}
    u\propto r^{\delta},\quad c\propto u,\quad \rho\propto r^{-\omega}
\end{equation}
(Note, that this form of the flow fields applies for $r>r_0$ where $r_0$ is the finite radius approached by $\xi_c R$ as $R$ diverges, see eq.~(\ref{eq:C+})). Since the distribution of flow fields at $t=0$ has no characteristic scale, the flow at $t>0$ will be described by a self-similar solution.

The $t>0$ self-similar solution may be constructed as follows. Denoting the scaling radius of this solution by $\tilde{R}$, we choose
\begin{equation}\label{eq:R_tilde}
    \dot{\tilde{R}}=-A\tilde{R}^\delta,
\end{equation}
with the same values of $A$ and $\delta$ obtained for the $t<0$ solution. This implies $\tilde{R}\propto t^{1/(1-\delta})$, i.e. $\tilde{R}$ is decreasing with $t$ and diverges as $t\rightarrow0_+$. Next, we choose a $C(U)$ curve that approaches in the limit $\tilde{\xi}\rightarrow0$, where $\tilde{\xi}\equiv r/\tilde{R}$, the same singular point approached in this limit by the $t<0$ solution, $U=C=0$:
\begin{equation}\label{eq:t0_prof}
    U=-U_1\tilde{\xi}^{\delta-1},\quad C=C_1 \tilde{\xi}^{\delta-1},\quad G=G_1\tilde{\xi}^{-\omega},
\end{equation}
with the same values of $U_1,C_1,G_1$ obtained for the $t<0$ solution (Note, that for $\delta>1$ any straight line $C\propto U$ is a solution of eq.~(\ref{eq:dUdC}) near $U=C=0$). It is straight forward to verify that the $t>0$ self-similar flow thus defined approaches the form given by eq.~(\ref{eq:t0}) as $t\rightarrow0_+$.

In the limit $t\rightarrow\infty$, $\tilde{R}\rightarrow0$ and the spatial distribution of the flow fields is described by the $\tilde{\xi}\rightarrow\infty$ behavior of the solution. In the limit $\tilde{\xi}\rightarrow\infty$, the $C(U)$ curve of the $t>0$ solution approaches the singular point $\{U=1-\delta,C=0\}$. The behavior of the self-similar solutions in the vicinity of this point was analyzed in \citep{WaxmanShvarts93}.

\section{Summary}
\label{sec:summary}

We have discussed the asymptotic behavior of adiabatic one dimensional flows of an ideal gas, which take place over a  characteristic scale $R$ that diverges or tends to zero. We have shown in \S~\ref{sec:general} that requiring the asymptotic flow to be independent of characteristic length scales implies, based on dimensional arguments, that the flow must be self-similar, with flow fields given by eq.~(\ref{eq:ss_scaling}) and $R$ satisfying eq.~(\ref{eq:Rdot}), $\dot{R}\propto R^\delta$. The ordinary differential equations determining the self-similar profiles are given by eqs.~(\ref{eq:dUdC})--(\ref{eq:deltas}).

In \S~\ref{sec:blast} we have shown that the asymptotic self-similar solutions describing the propagation of accelerating blast waves, propagating in a cold gas sphere of initial density $\rho\propto r^{-\omega}$ with $\omega>3$, are of the general form $\dot{R}\propto R^\delta$, with exponential solutions obtained at $\omega=\omega_c(\gamma)$. $\delta(\omega,\gamma)$ is shown in fig.~(\ref{fig:delta}) for $\gamma=4/3,5/3$. Fig.~\ref{fig:UC} demonstrates that numerical solutions of the hydrodynamic equations, eq.~(\ref{eq:hydro_eq}), indeed approach these self-similar solutions as $R$ diverges. The properties of the self-similar solutions obtained for $\omega\ge\omega_c(\gamma)$ are analyzed in \S~\ref{sec:WS}. The $C(U)$ curves of these solutions approach the singular point $\{U=0,C=0\}$ as $\xi\rightarrow0$. Analyzing the behavior of the solutions near this singular point we have shown that the energy and mass contained in region of the $\{\xi,R\}$ plane bounded by $\xi=1$ and $\xi=\xi_+(R)$, where $\xi_+(R)$ is a $C_+$ characteristic that approaches $\xi=0$ as $R$ diverges, both tend to finite constants as $R$ diverges.

For $\delta>1$, the shock radius diverges at a finite time, $t=0$. This implies that the spatial distribution of the flow fields at the divergence time is described by the $\xi\rightarrow0$ behavior of the solution, which is given by eq.~(\ref{eq:asym_dg1}). This implies that at $t=0$ the spatial distribution of the flow fields is given by eq.~(\ref{eq:t0}). In \S~\ref{sec:t0} we have shown that the $t>0$ flow is described in this case by a self-similar solution with the same value of $\delta$ as that of the $t<0$ solution, see eq.~(\ref{eq:R_tilde}), and spatial profiles determine by eq.~(\ref{eq:t0_prof}).

\acknowledgements We thank D. Kushnir for carrying out the numerical calculations presented in figure~\ref{fig:UC}. This research was supported in part by ISF, AEC, and Minerva grants.


\end{document}